\documentclass[aps,pra,twocolumn,superscriptaddress]{revtex4-1}

\usepackage{amsmath}
\usepackage{graphicx}
\usepackage{dcolumn}
\usepackage[colorlinks=true, allcolors=blue]{hyperref}
\usepackage{rotating}
\usepackage[super]{nth}
\usepackage{longtable}
\usepackage[colorinlistoftodos]{todonotes}

\begin{document}

\title{Isotope shifts in $^{20,22}$Ne - Precision measurements and global analysis in the framework of intermediate coupling}

\author{B. Ohayon}
 \email{ben.ohayon@mail.huji.ac.il}
\author{H. Rahangdale}

\affiliation{
 Racah Institute of Physics, Hebrew University, Jerusalem 91904, Israel
}
\author{A. J. Geddes}
\author{J. C. Berengut}
\affiliation{
School of Physics, University of New South Wales, Sydney, New South Wales 2052, Australia
}
\author{G. Ron}

\affiliation{
 Racah Institute of Physics, Hebrew University, Jerusalem 91904, Israel
}

\date{\today}

\begin{abstract}
We report new precision measurements of the $^{20}$Ne--$^{22}$Ne isotope shift for several transitions, as well as state-of-the-art, \textit{ab initio} field-shift calculations.
Our results are combined with historical measurements in a global fit to obtain the isotope shifts of all fifty low-lying neon levels with high precision. These level shifts show a wealth of electronic, nuclear, and relativistic phenomena.
Relying on the analogy between mass shift and fine-structure operators, we explain this plethora of neon level-shifts utilizing a small number of effective parameters in a global parametric investigation.
This investigation provides a birds-eye view on the isotope shift phenomena in noble gasses. From this vantage point, we reinterpret every effort made to calculate neon mass-shifts  \textit{ab initio}, and show that a remarkable agreement between experiment and theory is obtained.

\end{abstract}


\maketitle

\section{Introduction}

Upon analyzing the mass spectrum of neon ions emerging from canal rays, J. J. Thompson and W. Wein  observed a weak line at $22$ atomic mass units \cite{1912Thompson}. This observation led to the discovery that a specific element may be found with different mass, an isotope. Soon after, T. R. Merton suggested the possibility of detection of neon isotopes by spectroscopic means \cite{1921-Merton}, owing to small lineshifts between them generally referred to as isotope shifts (IS).

Isotope shifts  are minute differences between the electronic energy levels of  different isotopes of the same element. Today they are the focus of a multitude of theoretical and experimental efforts due to their importance for atomic and nuclear physics.  For few electron-systems such as hydrogen \cite{1998-HDradii,2011-HDradii}, helium \cite{1994-HeliumBreakthrough,2015He,2017-He}, lithium \cite{2000-LiTheory,2011Li}, and ions such as Ar$^{13+}$ \cite{2003-Ar, 2006-HCI}, accurate theory is available and so very precise measurements determine parameters such as charge radii \cite{2013-LuRMP}, as well as test bound-state QED \cite{2015-QED}.
The theory for isotope shifts of multielectron atoms is more involved; nevertheless, their IS are of interest for astrophysical searches for $\alpha$-variation \cite{2004-AlphaPRL,2007Koz,2011-AlphaVar}, probing for atomic Higgs and new Higgs-Like forces \cite{2017-Higgs}, as well as other new-physics scenarios \cite{2017-BSM}.

From a nuclear physics perspective, isotope shifts combine different precise atomic physics probes for studying nuclear properties, and their main use is nuclear-model-independent determination of (RMS) charge-radii differences \cite{1958-Brix,1989-Otten,2010-NuclearEffects,2016-Pearson}. The shape of some light isotopes ($A < 30$) vary wildly from the liquid-drop model \cite{2016-Pearson,2018-Clustering}, and display exotic phenomena such as proton and neutron halos \cite{1996-Halo,2004-Wang,2006-Li,2007-Muller,2009-Be}. On the other hand, the nuclear shape effect on the isotope shifts of light atoms is small compared with the total IS. Thus, for the determination of radii differences for light multielectron isotopes, both experiment and theory must display high precision and accuracy.

The isotope shifts of neon constitute a compelling case, and are the focus of this work. From the experimental perspective, a great multitude of lineshifts in the $^{20,22}$Ne pair has been measured over the last hundred years (see references in tables  \ref{tab:Lineshifts1}, and \ref{tab:Lineshifts2}), and including this work, some very precise results exist for a number of lines.  From a theoretical perspective, even-even neon isotopes, which do not possess a hyperfine structure, provide a clean environment to investigate the IS phenomena in light atomic systems with strong electron-hole interactions.

When analyzed globally, the plethora of experimental information gathered here not only gives an improved precision and accuracy in IS determination, but enables benchmarking and cross-checking theoretical calculations from a birds-eye perspective. We show that alleged large discrepancies between experiment and theory, which dominated theoretical work on neon isotope shifts in the last hundred years, are removed upon a reanalysis and reinterpretation enabled by our investigation.

The remainder of this paper is organized as follows. In Sec. \ref{chap:back} we provide a short background on the origins of the isotope shift phenomena and introduce the notations to be used in the following sections.

Sec. \ref{chap:FS} presents new, \textit{ab initio}, field shift  calculations, covering all relevant neon configurations. An immediate application of this calculation is a considerable update to the RMS charge radii differences of stable and unstable neon isotopes, and their uncertainty. The updated field shifts are used in the following sections for comparison between experiment and theory, since they were not taken into account in most calculations.

In Sec. \ref{Chap:Results} we present precision measurements of the $^{20,22}$Ne  isotope shifts for several transitions between the $2p^53s$ and $2p^53p$ manifolds. The small experimental uncertainty results from the use of the new measurement scheme of Dual-sideband Saturated absorption Spectroscopy \cite{2017-BGG}.
In Sec. \ref{chap:Transitions}, our measurements, as well as every other relevant isotope shift measurement existing in the literature, are averaged line by line, accounting for outliers and inconsistencies.

In Sec. \ref{chap:Levels}, we extract the residual-shifts of levels from a global linear fit to the average lineshifts. The lineshifts are then recalculated from levels resulting in a large increase in precision. The calculated lineshifts are presented in tables \ref{tab:Lineshifts1} and \ref{tab:Lineshifts2} and do not depend on any atomic physics theoretical framework.

In order to interpret, as well as significantly reduce the uncertainty in the level shifts we utilize the strong analogy between specific mass shifts (SMS) and fine-structure operators. First, in Sec. \ref{chap:IC}, we investigate the fine-structure of neon  in the framework of intermediate coupling to obtain an approximate expansion of the relevant wavefunctions over LS and \textit{jj}-coupled bases. Utilizing the angular coefficients of this expansion, we introduce in Sec. \ref{chap:Intra}, a small set of effective parameters for each configuration, with the same angular coefficients. These intraconfiguration parameters identify and differentiate between relativistic, \nth{1}, and \nth{2}-order SMS effects, and are extracted from fitting the average lineshifts of Sec. \ref{chap:Transitions} directly. Recalculating the level shifts from a smaller number of IS parameters dramatically reduces the uncertainty and so reveals the structure of poorly measured configurations.

In Sec. \ref{chap:Inter}, a smaller set of effective IS parameters is used in a similar way, relying on a deeper analogy between fine-structure and SMS effects. Thus a phenomenological understanding as to the origin of various SMS effects is gained,  and a greater precision in the levelshifts is obtained. 

In Sec. \ref{chap:Thoery}, we review all previous efforts for calculating neon isotope shifts \textit{ab initio}. Each theoretical effort is examined and reinterpreted in light of the phenomenological parameters of Sec \ref{chap:Intra} and  \ref{chap:Inter}, as well as the field shifts of Sec. \ref{chap:FS}. 

Section \ref{sec:out} is devoted to conclusions and outlook.

\section{Origins of the isotope shift}\label{chap:back}

Isotope shifts result from a change in the mass and the charge distribution of the nucleus, which corresponds to a mass shift (MS), which predominates in light elements;  and field shift (FS), which is the main effect in heavy atoms.  For a multi-electron-system, and in the non-relativistic limit, the mass shift is composed of a normal mass shift (NMS), resulting from a change in the reduced mass of the electron, and a specific mass shift (SMS), which is sensitive to electron correlations \cite{1930HughesEckart}.

Denoting the heavier isotope by $H$ and the lighter by $L$, the isotope shift of a transition with frequency $\nu$ is 
\begin{equation}
\delta \nu^\mathrm{L,H}=\nu_\mathrm{H}-\nu_\mathrm{L}
\end{equation}
and the NMS in the non-relativistic limit is defined as \cite{2013King}:
\begin{equation} \label{NMS}
\delta \nu^\mathrm{L,H}_\mathrm{NMS}=\nu_\mathrm{L} m\frac{M_\mathrm{H}-M_\mathrm{L}}{M_\mathrm{L}(M_\mathrm{H}+m)},
\end{equation}
where $m$ is the electron mass, and we adopt the convention that an isotope shift is positive when the line of the heavier isotope corresponds to the higher frequency.

The NMS as defined by (\ref{NMS}) is the main contribution to the IS of light atoms \cite{2013King}, and amounts to approximately $1$ GHz in the optical lines of neon; however, owing to precise neon mass measurements \cite{2008Mass}, one may subtract it without introducing appreciable contributions to the uncertainty. The largest relevant NMS uncertainty is in the most energetic lines ($63$ nm) and amounts to only a few kHz resulting from uncertainty in the wavelength \cite{2004Saloman}. 

 As early as 1930, Nagaoka and Mishima \cite{1930NM} observed a strong deviation from the NMS formula for neon lines, indicating there are other important effects. After subtracting the NMS, the remaining shift is denoted the residual isotope shift (RIS) \cite{1976BC}. It includes the SMS, the FS, and relativistic corrections to all effects, which are non-negligible in the level of precision presented here.

The SMS depends on correlations between electrons and its calculation is very sensitive to the details of the wavefunctions used \cite{2013King}.  In light elements ($Z<30$), it is by far the strongest contribution to the RIS for transitions and can be either positive or negative depending on the nature of correlations  \cite{1976BC}. It can be reliably calculated with very high precision only in very light elements with few electrons, five electrons being the forefront of such calculations \cite{2015-Boron, 2019-Bo}.

The FS in neon can only be observed with high precision, and so until the turn of the century have been considered negligible \cite{2000-Geit}. Since it is small, a \nth{1}-order picture is adopted here, where the change in energy of a level is proportional to the change in total electron probability density at the origin, times the mean-square charge radius difference $\delta\langle r^2 \rangle$ \cite{1932-FS}:
\begin{equation} \label{eq:FS}
\delta \nu_\mathrm{FS}=F\times\delta\langle r^2 \rangle,
\end{equation}
where $F$  the so called field-shift factor (FSF) of the level. Under the above convention, for transitions where the upper level has a lower electron density at the origin than the lower level, then the FSF is negative \cite{2012-NazeThesis}.

\section{Field-Shift factors and neon charge radii}\label{chap:FS}

\begin{table}[htbp]
  \centering
  \caption{Calculated field-shift factors of neon levels relative to the ground state, in MHz/fm$^2$. Uncertainties are estimated from the rate of convergence (parenthesis), and are dominated by uncertainty in the FS-factor of the ground state (square brackets). The last line is the field shift of the $614$ nm transition. Comparison with previous calculations is presented where available. The last column gives the field shift obtained from eq. \ref{eq:FS} utilizing radii measurements, with correlated uncertainty in square brackets dominated by nuclear model.}
  \begin{ruledtabular}
    \begin{tabular}{cccccr}
         & This Work      & \cite{2016-Wang}   & \cite{2012-NazeThesis}   & \cite{2000-Geit} & FS (MHz) \\\\
Method:  & CI+MBPT      & MCDHF  & MCDHF  &     GFS &  \\\\
   $ 3s$   & 142(2)[6]     & 138    & 139    &   & -44[5]   \\
   $ 3p$   & 112(1)[6]    &        &        &   & -35[4]   \\
   $ 4s$   & 118(1)[6]    &        &        &   & -37[4]   \\
   $ 3d$   & 114(1)[6]    &        &        &   & -35[4]    \\
   $ 4p$   & 111(2)[6]     &        &        &   & -35[4]    \\
   $ 5s$   & 115(2)[6]     &        &        &   & -36[4]    \\
   $ 4d$   & 112(1)[6]     &        &        &   & -35[4]    \\
   $ 4f$   & 112(1)[6]     &        &        &   & -35[4]    \\\\
    $3s [3/2]_2 $\\    $\rightarrow 3p [3/2]_2$ & -30.5(1.1)   & -32.3 & & -40(4)& 9.5[1.0] \\
    \end{tabular}%
    \end{ruledtabular}
  \label{tab:FS}%
\end{table}%

Whereas SMS effects are considered difficult to calculate \textit{ab initio} with high precision, calculations of the FSF of Eq. \ref{eq:FS} have reached the point where they are considered trustworthy in a multitude of scenarios, with uncertainties on the order of a few percent \cite{2015Marinova,2016-Pearson}.

We performed relativistic \textit{ab initio} field shift calculations in neon using the AMBiT software \cite{Kahl2019}. AMBiT calculates the electronic structure of a given atom using a combination of particle-hole configuration interaction (CI) and many-body perturbation theory (MBPT); this has been thoroughly detailed in \cite{Kahl2019, BerengutHgHoles2016}. 
The orbitals in a CI+MBPT calculation are divided into valence and core orbitals. The valence-valence correlations are treated with CI while MBPT is applied to the core-valence interactions \cite{DzubaCIMBPT1996}.

Our calculations in neon begin with a Dirac-Hartree-Fock (DHF) calculation. Each electron is treated as a single electron in a mean field $V^{N_\mathrm{DF}}$ arising from all electrons $N_\mathrm{DF}$ included in the DHF method. We selected a $V^{N-1}$ potential, which corresponds to the average potential from all but one of the electrons in neon. 
A single electron basis set is then constructed from B-splines in the DHF potential, these are used for the valence and virtual orbitals \cite{shabaev04prl,johnson88pra}.

The emu CI method, which is explained in detail in \cite{ageddesTaDb}, is used instead of conventional CI in order to decrease computational time and improve convergence. We construct our configuration state functions (CSFs) by single or double excitation of electrons and holes from the leading configurations $0$ (i.e. the closed-shell configuration $1s^2\ 2s^2\ 2p^6$), $2p^{-1} 3s^{1}$, $2p^{-1} 3p^1$ and $2p^{-1} 3d^1$. The '$-1$' superscript in the aforementioned configurations represents a hole state in a given orbital relative to the closed-shell configuration.
In neon, we restrict our single particle basis set to $13spdf$ for the CI calculation (that is, we only include excitations to valence and virtual orbitals up to $n = 13$ and $l = 3$). We also allow hole excitations to the $2s$ and $2p$ core shells, while correlations with the $1s$ orbital are included using MBPT. Our use of emu CI ignores correlations between high-energy doubly-excited states.
The MBPT method introduced in \cite{DzubaCIMBPT1996} has been implemented in AMBiT using the diagrammatical technique described in \cite{BerengutCI2006}. The neon calculation included one, two and three body core-valence diagrams at the second order of perturbation theory using a virtual basis set of $30spdfg$.

The Coulomb potential in AMBiT includes the effect of a finite nuclear charge distribution with a Fermi-Dirac distribution. Therefore we can use the ``finite field'' method to obtain field shift coefficients for neon. We modify the root-mean nuclear charge radii $\langle r^2 \rangle$ in regular increments, performing the entire CI+MBPT calculation each time to obtain transition energies $\omega$. The FSFs are extracted as
\begin{equation}
F=\frac{\delta \omega}{\delta \langle r^2 \rangle}.
\end{equation}

The results of our calculation for the FSF of each configuration, relative to the ground state, are given in table \ref{tab:FS}. The uncertainty is estimated from the convergence rate with a varying number of basis functions, as well as the ability of the code to recreate the level energies ($4\%)$. It is dominated by the ground state calculation. The FSF variance within each configuration is smaller than our quoted uncertainty, and thus for the current level of precision, we regard each configuration as having the same FSF, nevertheless, for specific transitions, the exact FSF is quoted. For the relative FSF between configurations, i.e. the 614 nm transition of which the IS was measured for unstable isotopes \cite{2011Marinova}, the uncertainty is slightly lower, as the ground state is not involved. We note that for non-s states, the FSF quickly converges to the ionization limit, estimated as $112(1)[6]$ MHz/fm$^2$.

The few \textit{ab initio} calculations that are available in the literature \cite{2016-Wang,2012-NazeThesis} are tabulated as well, and an agreement is found to the level of a few percent. Note that we reversed the sign of the field shift factor as it appears in \cite{2016-Wang}; when adding two neutrons to  $^{20}$Ne, the nuclear charge radius shrinks, resulting in a larger binding energy for the valence $2p^53s$ electron and a negligible field change for the $2p^53p$ electron. Thus $s-p$  transitions are more energetic for $^{22}$Ne, resulting in a positive field shift. 
Assuming that the calculations of \cite{2016-Wang}, which used the Multiconfiguration Dirac-Hartree-Fock method, have a comparable uncertainty, as indicated by his convergence rate, our results are in agreement.
Focusing on the $614$ nm transition, we consider a weighted FS-factor value of: $-31.4(0.9)$ MHz/fm$^2$. This value is four times as precise as, and disagrees with, the value of $-40(4)$ MHz/fm$^2$ estimated by \cite{2000-Geit} utilizing the semi-empirical GFS formula. It is noteworthy to mention that a 
 $25\%$ disagreement between \textit{ab initio} and semi-empirical calculations was found in Mg $s-p$ transitions \cite{2003-BerCalcs}.

Substituting the FSF of table \ref{tab:FS} in Eq. \ref{eq:FS}, along with the independently measured muonic X-ray measurements of $\delta\langle r^2 \rangle^{20,22}=-0.31[3]$ \cite{1995-Fricke}, we obtain the FS for each transition, with a correlated systematic uncertainty dominated by the nuclear model \cite{2011Marinova}.

We utilize the $614$ nm transition field shift to update the RMS charge radii differences of $^{17-19,21-26,28}$Ne, $\delta \langle r^2 \rangle^{20,A}$, utilizing the IS results presented in \cite{2011Marinova}. Given in table \ref{tab:RMS}, the effect of the updated FS-factor and its reduced uncertainty on $\delta \langle r^2 \rangle^{20,A}$, is dramatic. Eq. \ref{eq:FS} predicts that once a FS is given, there exists a linear relationship between the FS factor and the deduced radii differences; and indeed, the statistical uncertainty is larger by $25\%$; however, the updated values differ substantially from the previous. This difference stems from the fact that the FS-factor is used to calibrate the mass shift in a modified king-plot procedure \cite{2011Marinova}. We find a new mass shift factor $363.73(23)$ GHz u with uncertainty dominated by statistics, where the previous one $363.07[43]$ GHz u with uncertainty dominated by that of the FSF.

The updated $\delta \langle r^2 \rangle^{20,A}$ of table \ref{tab:RMS} posses a smaller total uncertainty and agree better with the x-ray measurements for stable isotopes \cite{1995-Fricke}, with the Droplet-model predictions of \cite{2011Marinova}, and with recent \textit{ab initio} coupled cluster calculations \cite{2016-SD}.

On top of updating the RMS charge radii differences, the FSF results presented in this section are utilised in Sec. \ref{chap:Thoery} for a comparison between analyzed neon isotope shift measurements and revisited mass shift calculations.

\begin{table}[htbp]
  \centering
  \caption{RMS charge radii difference in fm$^2$ with updated field shift factor from this work. Statistical uncertainty resulting from original IS measurements is in parenthesis. Correlated systematic uncertainty in square brackets and affect the general slope around the $^{20}$Ne point \cite{2015Marinova}.}
   \begin{ruledtabular}
    \begin{tabular}{lrrl}
    A     &\multicolumn{1}{c}{$\delta \langle r^2 \rangle^{20,A}$ \cite{2011Marinova}} & \multicolumn{1}{c}{$\delta \langle r^2 \rangle^{20,A}$ Updated}  &  \multicolumn{1}{c}{$r_\mathrm{ch}$ (fm)}\\\\
    
    17    &  0.220(29) [123] &  0.097(37) [092]& 3.022(6) [15]     \\
    18    & -0.207(15) [112] & -0.380(20) [071]& 2.942(3) [12]     \\
    19    &  0.017(19) [041] & -0.033(24) [029]& 3.001(4) [05]    \\
    20    &  0            &  0            &      3.006~~~~~[02]         \\
    21    & -0.217(14) [024] & -0.227(18) [022]& 2.968(3) [04]    \\
    22    & -0.321(04) [043] & -0.314(05) [040]& 2.953(1) [07]   \\
    23    & -0.571(34) [064] & -0.592(44) [059]& 2.906(8) [10]   \\
    24    & -0.627(19) [075] & -0.624(24) [072]& 2.900(4) [12] \\
    25    & -0.429(16) [122] & -0.336(21) [099]& 2.950(4) [17]  \\
    26    & -0.484(18) [143] & -0.374(22) [114]& 2.943(4) [19]  \\
    28    & -0.239(35) [213] & -0.004(44) [155]& 3.005(7) [26]   \\
    \end{tabular}%
    \end{ruledtabular}
  \label{tab:RMS}%
\end{table}%

\section{Precision Measurements of neon isotope shifts}\label{Chap:Results}

We use the method of Dual-sideband Saturated Absorption Spectroscopy (DSAS) to obtain the isotopic shifts between even neon isotopes with masses ${20}$ and $22$, for several optical transitions between the $2p^5 3s$ and $2p^53p$ configurations. In general, isotopes with an even number of nucleons do not possess a nuclear magnetic moment, which would lead to hyperfine structure that complicates both the measurement and interpretation of isotope shifts \cite{2013King}.

Briefly, the DSAS method \cite{2017-BGG}, utilizes two electro-optic-modulators (EOMs), each modulating a different laser beam split from the same source.  A high-frequency resonant EOM (here QUBIG EO-T1650M3 $1.5-1.7$ GHz) is introduced before a standard saturated absorption setup, and in practice folds the apparent isotope shift from roughly $1.6$ GHz to $50$ MHz, greatly decreasing the scan range of the laser to a few linewidths. A low-frequency EOM is placed at the entrance to a high-finesse Fabri-P\'erot (FP) cavity to create three narrow frequency-markers separated by the low EOM frequency. This signal acts as a calibrated `ruler'. By setting the low-frequency EOM driving frequency and amplitude, and FP offset so that two FP peaks merge with the folded IS peaks, the effect of laser frequency drift and scan nonlinearity is canceled to first order.
The experimental system is described in \cite{2017-BGG}, where the main difference in this work is that we use a narrow-band ($<$MHz) dye-laser in lieu of a home-built external cavity diode laser, enabling us to measure a variety of transitions in the optical regime with smaller inhomogeneous line-broadening. 

In \cite{2017-BGG}, systematic shifts resulting from pressure of $100$ mTorr, and RF- and laser-power were determined to be negligible ($<10$ kHz), and since the pressure shift is comparable for all of our measured lines \cite{1980-588-Press,1996-PressRev}, the pressure isotope shift is negligible at the current level of precision. Nevertheless, due to the variety of transitions inspected  in this work, we were able to determine the current limits of the DSAS method. We found two main systematic effects not accounted for in \cite{2017-BGG}:

The first is a slight asymmetry in the ruler signal, which is somewhat mitigated by removing the mode-matching lens before the FP and by filtering out high order harmonics from the LF EOM driver. The influence of this asymmetry on the IS uncertainty was accounted for by alternating the FP peaks between center and red-sideband, to center and blue-sideband, averaging each group separately, and taking the maximal difference between groups as systematic uncertainty $\sigma_\mathrm{FP}$.

The second effect comes from velocity changing collisions \cite{1988-VCC}, which contribute to the lineshape uncertainty in the fit to the IS signal, by introducing a broad gaussian background. By alternating the resonant EOM frequency between $\mathrm{IS}+50$ and $\mathrm{IS}-50$ MHz, thus switching the atomic signal peaks from $^{20}$Ne redshifted relative to $^{22}$Ne, to $^{20}$Ne blueshifted relative to  $^{22}$Ne, we dramatically change the background shape. The maximal difference between averages of each group is taken as lineshape uncertainty $\sigma_\mathrm{LS}$.

The total uncertainty is quoted conservatively as a linear sum of statistical and systematic uncertainties and results in $\sigma_\mathrm{tot}=0.1-0.2$ MHz giving a precision of at best $7\times 10^{-5}$. With the new systematics in mind, the $640$ nm isotope shift reported in  \cite{2017-BGG} was remeasured in a variety of conditions, and the recent value, along with all other measured transitions is reported in table \ref{tab:results}.

\begin{table}[htbp]
  \centering
  \small 
  \caption{Measured isotope shifts, $\delta \nu^{20,22}$ in this work, and uncertainty budget (MHz), for each wavelength in air (nm).
   $\sigma_\mathrm{stat}$ the statistical uncertainty, $\sigma_\mathrm{FP}$ the systematic uncertainty in the Fabri-P\'erot `ruler' signal,
   and $\sigma_\mathrm{LS}$ the lineshape uncertainty resulting mainly from assymetry induced by velocity changing collisions. The total uncertainty is taken as a linear sum.}
  \begin{ruledtabular}
    \begin{tabular}{cccccc}
    $\lambda$ (nm) & $\delta \nu^{20,22}$ (MHz) & $\sigma_\mathrm{stat}$ & $\sigma_\mathrm{FP}$  &$\sigma_\mathrm{LS}$ & $\sigma_\mathrm{Tot}$ \\
    \hline\\
    653.3       & 1588.238     & 0.019       & 0.081       & 0.117       & 0.22 \\
    650.6        & 1651.251     & 0.025       & 0.076       & 0.070       & 0.17 \\
    640.2       & 1626.051     & 0.014       & 0.046       & 0.050       & 0.11 \\
    638.3       & 1679.601     & 0.019       & 0.089       & 0.149       & 0.26 \\
    633.4       & 1641.052     & 0.016       & 0.053       & 0.091       & 0.16 \\
    630.5        & 1674.265     & 0.020       & 0.078       & 0.092       & 0.19 \\
    626.6       & 1647.375     & 0.013       & 0.115       & 0.016       & 0.14 \\
    \end{tabular}%
    \end{ruledtabular}
  \label{tab:results}%
\end{table}%

\section{Transition isotope shifts}\label{chap:Transitions}

For the global analysis in the following sections, every historical isotope shift in transitions between low-lying ($2p^54d$ and lower) levels, is useful.
We thus compiled and averaged the relevant lineshifts available in the literature, and presented them in tables \ref{tab:Lineshifts1} and \ref{tab:Lineshifts2} of the Appendix. These are roughly $260$ individual measurements reported in $40$ publications as early as 1930, spanning wavelengths of $60-8000$ nm.

For the lines for which $n\geq3$ measurements exist, we apply Chauvenet's criterion with a p-value of $0.05/n$ for identifying single outliers  \cite{1863-Chauvenet}. The six outliers found are italicized in table \ref{tab:Lineshifts2} and removed from the analysis. The fact that three of them were measured by the same group \cite{1992-KKR} may indicate an unaccounted-for systematic in that measurement campaign.  The weighted mean is recalculated after removal of outliers and appears in tables \ref{tab:Lineshifts1} and \ref{tab:Lineshifts2}.

To account for inconsistencies in the data, which were not removed along with the outliers, we inflate the standard uncertainty by $max(d_i,1)$, where  $d_i=\sqrt{\chi_i^2/\nu_i}$ is the Birge-Ratio for each line  \cite{1932-Birge}, and  $\nu_i$ the degrees of freedom (DOFs). In contrast with \cite{1997Basar}, we do not deflate the uncertainty of measurements which are in agreement. The line for which previous published results are most inconsistent is $3391.2$ nm with $d = 2.2$.

After the cleaning and averaging process we are left with  $145$ average transition isotope shifts which are in general more precise and accurate. For each line, we subtract the NMS using equation \ref{NMS}, to obtain the list of line RIS, denoted $\vec{T}$.
These lines connect $56$ levels.

\section{Residual Level Shifts}\label{chap:Levels}

The RIS of levels (denoted $\vec{L}$) is easier to interpret, from the perspective of atomic theory, than that of lines  \cite{1976BC,1997Pendrill,2008-Pendrill}. Moreover, due to the fact that there are many more lines than levels, recalculating the line IS from level IS results in a substantial increase in precision.

Since most lines connect to it, we chose the reference level, for which the RIS is held at zero, as $3s[3/2]_2$, and since this level is purely LS-coupled, we will refer to it as $3s$ $^3P_2$. This choice has no effect on any physical observables such as lineshifts \cite{1976BC}, but makes the covariance matrix (table \ref{tab:LevelCorr_2}) more diagonal. Owing to the independent and Gaussian nature of the average lineshifts, the level RIS and their uncertainties are readily obtained from a weighted multivariate linear regression \cite{2010-Glm}, to the following linear equation:
\begin{equation}
\vec{T} =\boldsymbol{ M_0} \vec{L},
\label{eq:LinesTrans}
\end{equation} 
where $M_0$ is a $145\times 55$ design matrix consisting of plus (minus) ones for upper (lower) levels, and with the reference level removed. The obtained level RIS and their uncertainties are presented in table \ref{tab:Level_RIS}. Retaining two significant digits in correlations, the correlation matrix separates to two blocks, shown in table \ref{tab:LevelCorr_2}. The obtained value $\chi^2/\nu=0.987$ for $\nu=90$ DOFs is in accordance with the most probable value for this distribution: $0.98\pm0.15$, and indicates the consistency of the analysis, as well as the applicability of the Birge-ratio uncertainty inflation employed in the averaging process of Sec. \ref{chap:Transitions}.

The IS of lines is recalculated from that of the levels and presented in tables \ref{tab:Lineshifts1} and \ref{tab:Lineshifts2}, with their uncertainties estimated utilizing the correlation matrix of table \ref{tab:LevelCorr_2}.  For most lines, our calculation yields results close to the historical average shift, with lower uncertainty. The average gain in uncertainty is $9$, with the highest gain of around $71$ for the poorly measured IS of the $625.9$ nm line. Lineshifts which have not been measured, including those belonging to forbidden transitions, and transitions involving the ground state, which are in the interest of recent investigation \cite{2015Sap,2017Sap}, can be calculated using tables \ref{tab:Level_RIS}, and \ref{tab:LevelCorr_2}.

Two lineshifts stand out since their calculated IS using the global fit is more than two standard deviations away from previous published results; these are the $783.9$ and $576.0$ nm lineshifts, which have only been measured by a single group \cite{1997Basar}, and so did not benefit from the historical averaging procedure. This observation demonstrates that the lineshifts calculated from the level shifts are less affected by unknown systematic errors, which may plague a single measurement, and so are expected to yield a higher accuracy.

In order to describe the level RIS using a small set of atomic physics parameters, we rely on the observation of Stone \cite{1959Stone}, that the operators contributing to the SMS, are analogous to those which contribute to the fine-structure. First we investigate neon energy levels to gain an understanding of the contributing factors and the compositions of each level.

\section{Intermediate coupling analysis of neon fine-structure}\label{chap:IC}

Excluding the neon ground state, which we denote $^1S_0$, all atomic neon electronic levels reported here are excitations of a single electron from the $1s^22s^22p^6$ configuration to principal and orbital quantum numbers $n$ and $l$, respectively. The $2p$ hole left by the excited electron results in a core angular momentum $j_\mathrm{c}$ of $1/2$ or $3/2$ creating two so-called subconfigurations. Adopting the definitions for the term $\vec{K}=\vec{j_\mathrm{c}}+\vec{l}$, and total angular momentum $\vec{J}=\vec{K}+\vec{s}$, all levels can be approximately identified using shorthand Racah notation: $nl[K]_J$ for the $j_\mathrm{c}=3/2$ subconfiguration, and $nl'[K]_J$, for the $j_\mathrm{c}=1/2$ subconfiguration. Since most excited states in neon do not conform to a particular coupling scheme, one should consider the above as simply a label, with only $J$ being a good quantum number.

The electron-hole interactions are described by LS-coupled basis functions, and evaluated using  Slater-Condon integrals: the Coulomb integrals $F_k(2p,nl)$, and the exchange integrals $G_k(2p,nl)$ \cite{1951-Spectra}. On the other hand, both the hole and the valence spins interact with their own orbits, to produce a fine-structure splitting $\zeta_{nl}$, described in a \textit{jj}-coupled basis. Thus the appropriate description scheme is intermediate-coupling (IC) \cite{1959Stone}, in which groups of levels with the same total angular momentum mix. Consequently, only the levels with $J=0,2$ of $ns$ sub-shells,  $J=0,4$ of level $nd$, and the $J=3$ level of $np$, are at the same time purely LS and purely \textit{jj}-coupled.

To obtain the compositions of each intermediate-coupled wavefunction in an LS basis, we follow the procedure outlined by \cite{1968-Cowan}. Throughout this section, the procedure is written explicitly for $p^5ns$ configurations, and is performed in an analogous way for the other configurations. First, the atomic Hamiltonian is approximated  by adding a diagonal Slater-Integral matrix, representing exchange and coulomb interactions in the central field approximation, and a spin-orbit block-matrix, where each block mixes different $J$s. For $p^5 ns$ configurations the IC matrix reads:
\begin{equation}
\label{IC}
\left(
\begin{array}{ccccc}
      & ^3P_2 &  ^3P_1 &  ^1P_1 &  ^3P_0 \\
^3P_2 & -G_1-\frac{\zeta_{2p} }{2} & 0 & 0 & 0 \\
^3P_1& 0 &  -G_1+\frac{\zeta_{2p}  }{2} & -\frac{\zeta_{2p} }{\sqrt{2}} & 0 \\
^1P_1& 0 & -\frac{\zeta_{2p}  }{\sqrt{2}} &  G_1 & 0 \\
^3P_0& 0 & 0 & 0 & -G_1+\zeta_{2p}  \\
\end{array}
\right)
\end{equation}
where we set the overall shift to zero by considering differences.

The \nth{1}-order Slater integrals, $G_1(2p,ns)$, account for the energy difference between $^3P$ and $^1P$ LS-states, and appear only in odd configurations. $\zeta_{2p}$  is the hole spin-orbit parameter which mixes $J=1$ states. The signs of fine-structure parameters which appear in \ref{IC} and its analogues for other configurations are chosen so that in a physical case, all parameters are positive.

The eigenvalues and eigenfunctions of the IC matrix represent the energies and compositions respectively of different levels, as a function of various fine-structure parameters; for $p^5 ns$ configurations they read:
\begin{widetext}
\begin{equation}
\begin{array}{cccc}
 -G_1-\frac{\zeta_{2p} }{2} & \zeta_{2p} -G_1 & \frac{1}{4} \left(\zeta_{2p} -\sqrt{16 G_1^2-8 \zeta  G_1+9 \zeta_{2p} ^2}\right) & \frac{1}{4} \left(\zeta_{2p} +\sqrt{16 G_1^2-8 \zeta_{2p}  G_1+9 \zeta_{2p} ^2}\right) \\
 \{1,0,0,0\} & \{0,0,0,1\} & \left\{0,\frac{4 G_1-\zeta_{2p} +\sqrt{16 G_1^2-8 \zeta_{2p}  G_1+9 \zeta_{2p} ^2}}{2 \sqrt{2} \zeta_{2p} },1,0\right\} & \left\{0,-\frac{-4 G_1+\zeta_{2p} +\sqrt{16 G_1^2-8 \zeta_{2p}  G_1+9 \zeta_{2p} ^2}}{2 \sqrt{2} \zeta_{2p} },1,0\right\}.
\end{array}
\label{eq:Eigensystem}
\end{equation}
\end{widetext}
Due to the simplicity of the eigenvalues of $p^5ns$ states, given in equation \ref{eq:Eigensystem}, the amount of mixing is described by a single parameter $\chi_n=(1+(4/3)G_1(n)/\zeta_{2p})^{-1}$ \cite{1981-Cowan}, which equals $0$ in the LS limit $(G_1>>\zeta_{2p})$, and $1$ in the \textit{jj} limit $(\zeta_{2p}>>G_1)$.  

To determine the values of fine-structure parameters for each configuration, we utilize the fact that neon energy levels are very well known \cite{2004Saloman}, and employ a standard nonlinear $\chi^2$-minimization procedure to reproduce them.
In accordance with \cite{1970-NeAlpha}, we find that for $np$ configurations, inclusion of the Trees-Racah effective operator  $\bar{\alpha}$ in the IC matrix \cite{1951-Trees,1952-Racah,1968-effectiveOperator}, with LS-coefficients $L(L+1)$, considerably improves the fit. This operator accounts for the effect of interaction with distant configurations. An inclusion of a valence spin orbit parameters $\zeta_{nl}$ was found to improve the fit for  $np$ levels but not for $nd$ levels, where it is vanishingly small.  The results are presented in table \ref{tab:FSpars}.

Inspecting the parameters of table \ref{tab:FSpars}, the best estimation for the fitting quality is that within uncertainty, the same $\zeta_{2p}$ values are found for each configuration, and are close to  $\zeta_{2p}=520.283(1)$ cm$^{-1}$, the core fine-structure splitting measured from the spectrum of Ne-II \cite{2006-Kramida}.

Except for the inclusion of $\bar{\alpha}$ in the above considerations, we neglected to explicitly include configuration interaction, which pushes interacting close energy levels apart, since it is negligible in all lower ($n\leq4$) configurations of neon \cite{1973Keller}, and in the $5s$ configuration \cite{1997Pendrill}. We note that appreciable configuration interaction is expected as levels coalesce towards the ionization limit. In neon it has been identified in the $6,8,10,11s$, $7,8p$, and $6,7,9d$ configurations \cite{1951-Spectra, 1997Pendrill}, and so the IS extrapolations to the ionization limit relying on these configurations reported in \cite{1994-74nm,2016-Sap}, should be examined.

Substituting the fine-structure values of table \ref{tab:FSpars} to the IC eigenvectors  (for which equation \ref{eq:Eigensystem} is an example)  for each configuration, results in the parametrization of neon wavefunctions over an LS-coupled basis. The parentage is obtained by squaring the eigenvector coefficients and is written in table \ref{tab:AC2}. The core parentage is obtained by expanding  each LS-term over the \textit{jj}-basis \cite{1981-Cowan}, and demonstrates that $p^5 4s$ and higher levels are practically \textit{jj}-coupled.

Armed with the angular coefficients for each term, the RIS of levels is explained by a small number of effective IS parameters, analogous to some of the fine-structure parameters of table \ref{tab:FSpars}.

\begin{table}[htbp]
\scriptsize 
  \centering
  \caption{ \label{tab:FSpars}
  Fine structure parameters, obtained from nonlinear fitting of  neon fine structure (cm$^{-1}$).
  Column heads denote the $nl$ configuration of the valence electron.
  $\zeta_{2p}$ is the $2p$-hole spin-orbit interaction. $F_2$ is a second-order Coulomb integral between valence electron and $2p$-hole where the zero order was set to zero by considering differences. The $G_n$ are the nth-order exchange integrals between valence and $2p$-hole. $\zeta_{nl}$ a spin-orbit interaction found to be non-negligible only in $np$-configurations. $\bar{\alpha}$ is the Racah-Trees effective parameter (see text) which accounts for small distant configuration interactions.
  }
  \begin{ruledtabular}
    \begin{tabular}{lrrrrrrrr}
                & \multicolumn{1}{c}{\textbf{3s}}   & \multicolumn{1}{c}{\textbf{3p}}& \multicolumn{1}{c}{\textbf{4s}}& \multicolumn{1}{c}{\textbf{5s}}& \multicolumn{1}{c}{\textbf{3d}}              & \multicolumn{1}{c}{\textbf{4p}}              & \multicolumn{1}{c}{\textbf{4d}}  \\
                \cline{2-9}\\
    $\zeta_{2p}$  & 518.0(3)        & 518(3)         & 519.0(4)        & 518.7(1)        & 520.1(8)        & 518(4)         & 519.8(8) \\
    $F_2$                         &             & 157.2(3)        &                &             & 16.3(1)        & 44.4(6)         & 6.7(1) \\
    $G_0$&                        & 766.3(6)        &             &               &             & 243(2)         &             &  \\
    $G_1$& 743.5(3)             &             & 174.9(6)        & 67.5(2)        & 3.0(2)        &             & 1.7(2) \\
    $G_2$&               & 37.2(5)        &             &             &             & 12(2)         &             &  \\
    $G_3$&             &             &      &        & 0.03(3)                &             & 0.00(4) \\
    $\zeta_{nl}$    &        & 8(2)         &             &            &       & 3(5)         &         &  \\
    $\bar{\alpha}$      &            & 31.7(7)        &             &                    &             & 8(2)          &  \\
    \end{tabular}%
    \end{ruledtabular}
\end{table}%

\section{Intraconfiguration fit}\label{chap:Intra}

Inspecting the level RIS of table \ref{tab:Level_RIS},  the first striking feature is a large difference between the centers of gravity of all different configurations, which is around $10$ GHz for the ground level, and up to $1$ GHz between other levels. Whereas most of this offset results from SMS effects, up to $44$ MHz are attributed to field-shift effects of table \ref{tab:FS}.

In the following paragraphs we introduce a series of effective intraconfiguration isotope shift parameters, analogous to a subset of the fine-structure parameters of table \ref{tab:FSpars}.
These parameters, together with the angular coefficients derived in Sec. \ref{chap:IC} and presented in table \ref{tab:AC}, describe the level RIS in an order-by-order basis.

First we denote the RIS offset of  $p^5nl$ $^3P$ terms from the $p^53s$ $^3P_2$  reference level, as $f_0(nl)$. This introduces $7$ RIS offset parameters, one for each configuration other than the reference. These offset parameters constitute all order SMS, the FS of table \ref{tab:FS}, and possible relativistic corrections which we deduce that are small outside of the $2p$-core.

 Due to the negligible field-shift differences within configurations, intraconfiguration effects are pure SMS.
 \nth{1}-order SMS 
 is analogous to the $G_1$ fine-structure parameter of table \ref{tab:FSpars}\cite{1959Stone}, which differentiates between $^1P$ and other LS-terms. We thus introduce  $5$  \nth{1}-order SMS parameters $g_1(2p,nl)$, one for each odd configuration, with the same angular coefficients as the $^1P$ terms of table \ref{tab:AC2} \cite{1976BC}. 

Since spin-orbit interaction is significant in the core, we introduce $7$ core-shift $z_{2p}$ parameters, one for each configuration other than the ground state, analogous to $\zeta_{2p}$ of table \ref{tab:FSpars}, with the same angular coefficients as the $j_\mathrm{c}=1/2$ parentage of table \ref{tab:AC2}. This $J$-dependence for pure LS-terms was first observed in neon \cite{1971BK}, and is a genuine signature of IS relativistic effects \cite{1976BC}, which for mass shift operators are order $(\alpha Z)^2$ \cite{2007Koz}. Including spin-orbit parameters for the valence electron was not found to improve the fit and we conclude that they are negligible in the current level of precision.

The origins for this ionic-core RIS has been ascribed to a few large ($\sim 100$ MHz) relativistic IS effects which partially cancel each other: The first is the relativistic spin-orbit interaction, a one-body operator which may be identified as the relativistic correction to the NMS. Two two-body relativistic corrections have been identified as well, the spin-other-orbit contribution, and Stone's term denoted $\Delta_1$ \cite{1963Nuclear}, which does not have a fine-structure analogue \cite{1985-VesFS}. The fourth relativistic correction, which is difficult to calculate, is the crossed second order CSO effect between the SMS operator and various magnetic terms in the Hamiltonian \cite{1973Keller}.

Except for the core shift, and excluding higher order terms, all $np$ levels are expected to have the same SMS, since no $G_1$ Slater integral appears in the energy difference between terms \cite{1959Stone}. The fact that $np$ levels do not have the same RIS was first evident in the measurements of \cite{1965Odin} but overlooked later in the analysis of \cite{1992-KKR,1985-Ahmad}. Since the LS-dependent relativistic corrections which appear in \cite{1961stone,1963Nuclear}, are estimated to be much smaller, this difference has been attributed to the CSO effect of the SMS and electrostatic energy operators \cite{1973Keller}. A brute-force approach, which we follow here, for accounting for \nth{2}-order SMS is to include one effective IS parameter for each LS-term \cite{1976BC}. For each of the two $np$ configurations, and since the $^3P$ shift is absorbed in $f_0$, we include 5 parameters $T(^{2S+1}L)$ with the same angular coefficients as the respective LS-terms in table \ref{tab:AC2}.

\begin{table}[!bp]
  \centering
  \footnotesize
  \caption{Effective residual isotope shift parameters from intraconfiguration global fit (MHz).
  $z_{2p}$ is the $2p$-hole relativistic shift parameter, analogous to the $\zeta_{2p}$ hole spin-orbit fine structure parameter of table \ref{tab:FSpars}.
  $f_0$ is the offset of the $^3P$ level of each configuration from the $p^53s$ $^3P_2$  reference level, resulting from all-order SMS and the FS denoted in table \ref{tab:FS}.
  $g_1$ is the \nth{1}-order SMS parameter differentiating $^1P$ and $^3P$ terms of odd configuration, analougous to the $G_1$ Slater-integrals of table \ref{tab:FSpars}.
  $T(^{2S+1}L)$ are \nth{2}-order SMS parameters for LS-terms, present only in the even configurations.
  $c$ is an effective configuration-interaction parameter which translates the $3p$'$[1/2]_1$ level.
  }
  \begin{ruledtabular}
    \begin{tabular}{lcccccc}
        & \textbf{3s}   & \textbf{3p}  & \textbf{3d} & \textbf{4p} & \textbf{5s}  & \textbf{4d}     \\ \cline{2-7}\\
 $z_{2p}$         & 23.7(2)               & 23.6(9)    & 27(2)  & 27(2)    & 22(2)                & 23(2) \\
 $f_0$&& 419.4(8)      & 241(3)    & 280(3)     & 186(2)     & 233(1) \\
 $g_1$& -726(1)&& -107(4)&& -64(5)& -58(7) \\
 $T(^1S)$    && -87(2)       && -31(2)& &  \\
 $T(^3S)$    && 130(2)       && 69(17)&&  \\
 $T(^1P)$      &                              & 65(2)        &            & 14(2)                         &                    &                      \\
 $T(^1D)$    &                              & 45(1)        &            & 5(3)                         &                    &                      \\
 $T(^3D)$    &                              & 39.0(7)      &              & 16(1)                         &                    &                      \\
$c$                 &             & 4.6(6)                                                      
    \end{tabular}%
    \end{ruledtabular}
  \label{tab:ParIntra}%
\end{table}%

The level RIS of the $3p$ configuration is known with very high accuracy, and so minute effects, which were not necessary to be taken into account in the analysis of \cite{1971BK}, are found to play a role here. Due to the small ($<1\%$) but known CI mixing of the most energetic $3p$ levels \cite{1973-Lib}, we find that an effective CI parameter has to be included for the two highest (and so most mixed) levels. The highest level is entirely $^1S$, and so CI shift is already absorbed in $T(^1S)$; we thus included a single effective CI parameter $c$ which translates the $3p$'$[1/2]_1$ level.

The $30$ \nth{1} and \nth{2}-order effective IS parameters are denoted $\vec{P}$ and follow a linear relation to the RIS of levels:
\begin{equation}
\vec{L} = \boldsymbol{M_1} \vec{P}
\label{eq:ParLines}
\end{equation}

Where $M_1$ is a $55\times 30$  block matrix of angular coefficients from table \ref{tab:AC2}, and each block corresponds to a configuration. In principle it is possible to employ Eq. \ref{eq:ParLines} for the levels of each configuration separately and obtain the isotope shift parameters, this procedure is often called a `parametric investigation' and is reviewed in \cite{1976BC}. Nevertheless, since the level shifts are derived from transition shifts, they are in some cases extremely correlated, and do not necessarily have a Gaussian distribution, a fact which may add complexity and ambiguity to the analysis.

A true and simple linear global fit, relying on uncorrelated measurements with normally distributed uncertainties, connects the theoretical parameters with the measured quantities directly. In practice, combining equations \ref{eq:LinesTrans} and \ref{eq:ParLines} results in a linear relationship between transitions and IS parameters:
\begin{equation}
\vec{T} = \boldsymbol{M_{01}} \vec{P}
\label{eq:ParTrans}
\end{equation}
where $M_{01}=M_0\times M_1$ a $145\times 30$ design matrix. The fitting value for $\nu=115$ DOFs was $\chi^2/\nu=0.937$, and the obtained parameters are presented in table \ref{tab:ParIntra}.

\begin{figure}[htbp]
\includegraphics[width=0.85\columnwidth, trim={135 200 90 195},clip]{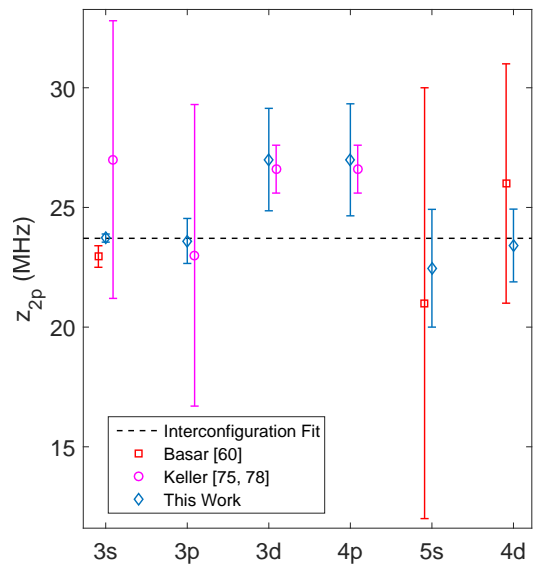}
\caption{\label{fig:Z}
Relativistic core fine-structure SMS (MHz) in various configurations. Compared with the values obtained by Basar \textit{et. al.} \cite{1997Basar}, and Keller \textit{et. al.} \cite{1973Keller,1971BK}. The dashed line notes the value returned from the interconfiguration global fit of Sec. \ref{chap:Inter}}.
\end{figure}

The level RIS are recalculated from the IS parameters and their correlations, and presented in table \ref{tab:Level_RIS}. The application of an intra-configuration fit yields an average gain in uncertainty for level RIS of $10$, revealing the structure of the $3d, 4p$ configurations. Calculating lineshifts from IS parameters and their correlations, the average gain in uncertainty over the historical average is $20$, demonstrating the strength of this technique. For the three $4s$ levels, there is an equal number of parameters, and so this configuration is dealt with in the next section. For all other valence-hole configurations, the similar values of the relativistic core shift $z_{2p}$ indicate the consistency of analysis; these are plotted in figure \ref{fig:Z}, where they are compared with previous determinations \cite{1971BK,1973Keller,1997Basar}. We note that the lower uncertainty reported by \cite{1973Keller} is an attribute of the statistical formalism used.

\section{Interconfiguration fit}\label{chap:Inter}

The similar values of $z_{2p}$ of figure \ref{fig:Z} indicate that combining them to a single parameter - the SMS in the Ne-II fine-structure, will improve the global fit. Another reduction of parameters is possible by taking the isotope-shift fine-structure analogy further, through assuming that the $g_1$ parameters of table \ref{tab:ParIntra} not only share the $G_1$ parameters' angular coefficients, but also decrease with the same proportion:
\begin{equation}
\frac{g_1(2p,nl)}{g_1(2p,n'l)}=\frac{G_1(2p,nl)}{G_1(2p,n'l)}.
\label{eq:g1}
\end{equation}

\begin{figure}[htbp]
  \centering
\includegraphics[width=\columnwidth, trim={8 190 120 180},clip]{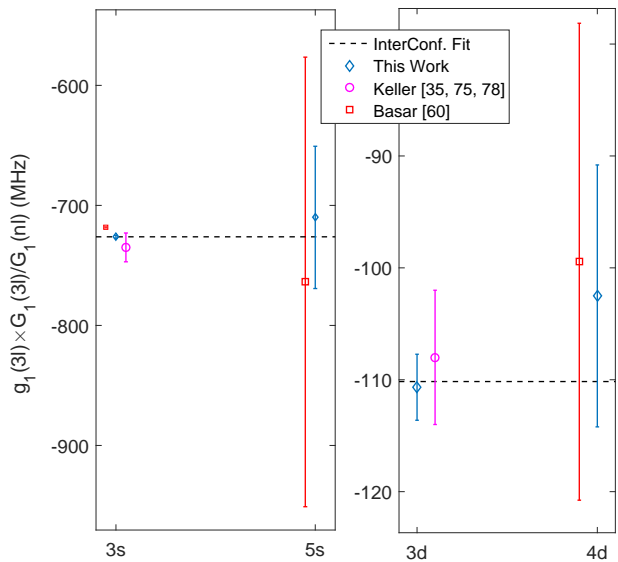}
\caption{\label{fig:g}
\nth{1}-order SMS parameters $g_1$ from table \ref{tab:ParIntra}, and those deduced from Baser \textit{et. al.} \cite{1997Basar}, and Keller \textit{et. al.} \cite{1971BK,1973Keller,1976BC}. $g_1(nl)$ is scaled to $g_1(3l)$ according to the respective fine-structure exchange integrals using equation \ref{eq:g1}. The dashed line notes the value returned from the interconfiguration global fit of Sec. \ref{chap:Inter}.}
\end{figure}

Figure \ref{fig:g} demonstrates the validity of equation \ref{eq:g1} for the  relevant configurations.  

The above discussion leads to a global fit with $21$ parameters (table \ref{tab:ParInter}), and $\nu=124$ DOFs, which gives $\chi^2/\nu=0.930$.  The value of the global core SMS parameter is shown in figure \ref{fig:Z}, and of the relevant $g_1(2p,3l)$ parameters in figure \ref{fig:g}. As in Sec. \ref{chap:Intra}, the level RIS are recalculated and given in table \ref{tab:Level_RIS}, and show an average gain of $2.5$ in uncertainty between intra- and inter-configuration fit. Recalculating transition shifts from the parameters of table \ref{tab:ParInter}, and their correlations, results in an average gain of $27$. 

The inter- and intraconfiguration effective RIS parameters of tables \ref{tab:ParIntra} and \ref{tab:ParInter} offer an informed method of comparing the analyzed experimental results with theoretical calculations, since the latter usually did not account for one or more non-negligible effects.

\begin{table}[htbp]
  \centering
  \footnotesize
  \caption{Effective IS parameters for inter-configuration global fit (MHz). 
  Compared with table \ref{tab:ParIntra}, the $z_{2p}$  parameters are fused together, and  some $g_1$ parameters are deduced from Eq. \ref{eq:g1}. 
  }
  \begin{ruledtabular}
    \begin{tabular}{lccccccc}
			& \textbf{3s}& \textbf{3p}& \textbf{4s}& \textbf{3d}& \textbf{4p}& \textbf{5s}& \textbf{4d}  \\ \cline{2-8} \\
   $z_{2p}$              &   23.7(2)            \\
     $f_0$                     &        					 & 419.4(7) & 144(2)     & 244(2)     & 283(3)     & 186(1)            & 233(7) \\
     $g_1$                                               & -726(1)   &							&						&-110(3)\\  
$    T(^1S)  $          &                    & -87(2)            &   &       & -32(2)                      \\
$    T(^3S)   $             &                & 130(2)            &    &      & 69(15)                           \\
$    T(^1P)    $            &                & 64(1)            &     &     & 14(2)         &    &                        \\
$    T(^1D)     $           &                & 45(1)            &     &     & 6(3)          &   &                                  \\
$    T(^3D)      $          &                & 39.0(7)            &   &       & 16(1)       &      &                                \\
$  c        	$	           &				& 4.6(6)               &                             
    \end{tabular}%
    \end{ruledtabular}
  \label{tab:ParInter}%
\end{table}%

\section{Comparison with theory}\label{chap:Thoery}

In this section we go over every theoretical effort made in the past to calculate neon isotope shift for the relevant transitions. We show that many large discrepancies are removed when reinterpreting the \textit{ab initio} calculations of specific mass shifts as calculations of the effective parameters of tables \ref{tab:ParIntra} and \ref{tab:ParInter}, while taking into account the field shifts of table \ref{tab:FS}.

The first effort for calculating SMS effects in isotope shifts was undertaken by Bartlet and Gibbons in $1933$ \cite{1933-BG}. They used Hartree-Fock (HF) wavefunctions to evaluate the isotope shifts in neon $3s-3p$  transitions measured by \cite{1930NM}. Since LS-coupling was assumed, and no FS, core shift, and CSO effects were taken into account, their results may now be reinterpreted as a calculation of \nth{1}-order SMS parameters. Adapting their notations to the conventions of this work, and utilizing modern values for the neon and electron masses, their resulting LS SMS parameter: $g_1(3s)=$ $3s$ $^1P$ - $3s$  $^3P$ = $-589$ MHz, is now in a reasonable agreement, considering the number of HF functions used \cite{1985-Ahmad}, with the value of $g_1(3s)=-726(1)$ MHz from table \ref{tab:ParInter}. 

More recently, an MBPT calculation relying on the same assumptions as \cite{1933-BG}, including LS coupling, was attempted by \cite{1985-Ahmad}. Thus it can not be directly compared with experiment in a meaningful way.  Taking the difference between their value for the $3s$ $^1P-3p$ SMS, and $3s$ $^3P-3p$ SMS to the first-order of the computation (which is the only order expected to contribute to this value) returns a value of $g_1(3s)=-720$ MHz, which is agreement with our analysis to $1\%$.
This agreement supports the fact that relativistic effects do not contribute substantially to SMS outside the core.

Other estimates of \nth{1}-order SMS parameters appear in the literature for parametric investigations in neon, and may be compared directly with the values appearing in tables \ref{tab:ParIntra} and \ref{tab:ParInter}. Our value of $g_1(3d)=-106(4)$ MHz, agrees with the value of $3d$ $^1P$ - $3d$  $^3P$ = $-96(6)$ MHz calculated by \cite{1973KL} using multiconfigurational Hartree-Fock (MCHF) computations \cite{1939-Hartree}, with uncertainty estimated from the difference between their results for two calculation methods. Our value of $g_1(4d)=-58(7)$ MHz agrees well with the value of $-58$ MHz computed by Bauche using the code of \cite{1970-Fischer}, as it appears in \cite{1979GBC}, however, our value of  $g_1(5s)=-64(5)$  MHz is in less agreement with the calculated value of $-54$ MHz appearing in  \cite{1979GBC}.

As for the offset between configurations, both \cite{1933-BG} and \cite{1985-Ahmad} calculated $3s-3p$ isotope shifts assuming no appreciable FS, which from table \ref{tab:FS} is $9.5(1.0)$ MHz, and negligible differences within the $3p$ configuration, where from table \ref{tab:Level_RIS}, the differences are up to $200$ MHz. 
For comparison purposes, we compute the center of gravity of the $3p$ configuration relative to the reference level, by weighing each LS-parameter $T(^{2S+1}L)$ of table \ref{tab:ParInter},
by its multiplicity $w_{SL}=(2S+1)(2L+1)$. The resulting center configuration relative RIS is $ \delta  \nu_\mathrm{RIS}(3p)=f_0(3p)+\Sigma_{SL\neq ^3P}w_{SL}T(^{2S+1}L)/\Sigma_{SL}w_{SL}=456(3)$ MHz. 
For comparing the SMS itself, we remove the relative field shift of $9.5$ MHz appreaing in table \ref{tab:FS} to obtain a value of $ \delta \nu_\mathrm{SMS}(3p)=447(3)$ MHz. This value is in reasonable agreement with the value of $3s$ $^3P$ $-3p$ $=410$ MHz of \cite{1933-BG}, and is in very good agreement with the value of  $3s$ $^3P$ -$3p$ $=447$ MHz calculated by \cite{1985-Ahmad}  to \nth{1}-order. \nth{2}-order SMS effects are not expected to contribute to the center of gravity shift. This agreement further validates that relativistic IS effects do not play a significant role outside of the core.

For relating the relativistic core fine-structure IS of our analysis with theory, the separation between NMS and SMS is not accurate, and the comparison is made by considering the total core fine-structure IS. Applying equation \ref{NMS} to the Ne-II fine-structure interval of $780.424(2)$ cm$^{-1}$ \cite{2006-Kramida}, we obtain a core fine-structure non-relativistic NMS of $58.35$ MHz, with negligible uncertainty. Adding to that the global SMS parameter $z_{2p}=23.72(16)$ from table \ref{tab:ParInter}, the total core fine-structure isotope shift is estimated as $82.07(16)$ MHz. This value compares well with a difficult, \nth{4}-order, MBPT calculation by Veseth, which under our sign convention results in $93(14)$ MHz, with the uncertainty evaluated as $10\%$ of the large CSO term in the calculation \cite{1985-VesFS}.

Since the optical electron is at rest at infinity, the RIS of all levels approaches zero with respect to the appropriate subconfiguration ionization limit. The $f_0$ values of table \ref{tab:ParInter} represent the RIS of pure $j_\mathrm{c}=3/2$,$^3P$ states, which for $ns,nd$ configurations is a non-relativistic, \nth{1}-order effect and expected to approach zero in an ordinary (polynomial) manner. From this convergence rate, a crude estimation of the absolute RIS of the reference level is $210(20)$ MHz, resulting in an absolute RIS of the ground level of $-9392(23)$ MHz. 
From the fast convergence of the FS-factors to the ionization limit (table \ref{tab:FS}), we assume that the absolute field shift of $4f$ state is negligible, and so estimate the absolute FS of the ground level by $\delta\nu_\mathrm{FS}$($^1S_0 )=-35(4)$ MHz, giving an absolute $\delta\nu_\mathrm{SMS}$($^1P_0)=-9358(23)$ MHz. This value agrees to some extent with the calculated value of $-10.1$ GHz calculated by \cite{1985-Veseth}, if we follow \cite{1994-74nm} in estimating the calculation uncertainty as $0.5$ GHz.

To conclude this section we review the only available direct calculation of various IS effects for a specific transition, namely the $614.3$ nm transition between our reference level $3s[3/2]_2$ and the  $3p[3/2]_2$ level \cite{2016-Wang}. This line isotope shift is of high interest for nuclear physics, since it was measured precisely for various neon isotopes, including the two-proton halo candidate, $^{17}$Ne \cite{2008-Radii,2011Marinova}, and determines the RMS charge radii difference between them. The calculation method used was multiconfigurational Dirac-Hartree-Fock (MCDHF), including Breit interactions and not including other relativistic corrections. A direct comparison with the shifts observed in experiment, for which the uncertainty was $1-3$ MHz, yielded discrepancies from $15$ to $97$ MHz (see table 4 in \cite{2016-Wang}).

\begin{table}[!bp]
  \centering
  \small
  \caption{Isotope shift effective parameters compared with theoretical calculations.
  The first lines reinterpret calculations of mass shifts within configurations as those of effective $g_1$ parameters from tables \ref{tab:ParIntra} and \ref{tab:ParInter}.
  The following lines reinterpret calculations of interconfiguration offsets utilizing the field shifts of table \ref{tab:FS}.
  All values are in MHz.}
  \begin{ruledtabular}
    \begin{tabular}{lcccc}
    Parameter          & This Work                & Theory            & Ref.               & Method \\ \cline{1-5} \\

     $g_1(3s)$             & -726(1)                  & -720               & \cite{1985-Ahmad}            & MBPT \\
                       &                    &           -591               & \cite{1933-BG}               & HF \\
    $g_1(3d)$             & -106(4)                  & -96               & \cite{1973Keller}         & MCHF \\
    $g_1(4d)$             & -58(7)                  & -58               & \cite{1979GBC}             & MCHF \\
    $g_1(5s)$             & -64(5)                  & -54               & \cite{1979GBC}            & MCHF \\\\

    $\delta\nu_\mathrm{SMS}(3s$ $^3P_2-3p)$         & 447(3)                          & 447                         & \cite{1985-Ahmad}               	  & MBPT	 \\
                       &                       & 410                         & \cite{1933-BG}                          & HF \\\\
                       
                       $\delta\nu$(fine structure)       & 82.1(2)                 & 93                & \cite{1985-VesFS} & MBPT \\\\
                       
    $\delta\nu_\mathrm{SMS}$(Core-Ionization limit)     & -9359(23)                 & -10110            & \cite{1985-Veseth} & MBPT \\\\
   $\delta\nu(3s$ $^3P_2-3p[3/2]_2)$    & 1663.8(2)                 & 1668            & \cite{2016-Wang} & MCDHF \\\\

    \end{tabular}%
    \end{ruledtabular}
  \label{tab:th}%
\end{table}%

These discrepancies are to a large extent removed upon reanalysis of the theory.  After reversing the sign of the field shifts of table 4 in \cite{2016-Wang}, the discrepancies are even larger for all isotopes, and vary between $30$ and $110$ MHz.
Another large correction is related to the Breit interaction. In this work we demonstrated that relativistic effects are mostly confined to the core, where the unaccounted-for relativistic corrections to the mass shift operators cancel the Breit interaction to a large extent \cite{1985-VesFS}. Moreover, the $3s[3/2]_2$ level is completely related to the $j_\mathrm{c}=3/2$ core, and the $3p[3/2]_2$ level is $89\%$ related to this core (see table \ref{tab:AC2}). We thus exchange the Breit correction of $9$ GHz u ($27$ MHz for $^{20,22}$Ne) of table 3 in \cite{2016-Wang}, with the much smaller value of $0.57$ GHz u, corresponding to a $11\%\times z(2p)=2.6$ MHz for $^{20,22}$Ne core shift. 
After application of both corrections, and assessing the theoretical uncertainty as at least the last presented digit, an agreement between experiment and theory is achieved for all neon isotopes. The corrected theoretical value for   $^{20,22}$Ne is $1668(5)$ and agrees with the value measured by \cite{2008-Radii} of $1663.6(1.7)$ MHz, and the updated value from this work of $1663.8(2)$.

An overview of table \ref{tab:th} demonstrates that when relativistic effects do not play a key role, i.e. outside of the core, MBPT and MCHF methods handle multielectron correlations much better than is most often claimed \cite{2015Marinova}, at least for the case of neon transitions.

\section{Conclusion and outlook}\label{sec:out}

\subsection{Neon charge radii}

In this work we calculated the field-shift factors for a variety of neon transitions with an accuracy of a few percent. These values are used in updating the RMS charge radii differences for a multitude of neon isotopes, as well as facilitating a comparison of mass shift parameters with theoretical calculations.

Following our update, the limiting factor for determining neon charge radii differences is the systematic uncertainty in the modified King-plot procedure stemming from the calibration of absolute $^{20-22}$Ne radii measurements. A better calibration of the slope is obtained provided that $\delta \langle r^2\rangle^{20,22}$ is known with higher precision. Whereas we elected to use the same, nuclear-model dependent, value for $\delta \langle r^2\rangle^{20,22}$ that is used by \cite{2008-Radii,2011Marinova}, other determinations appear in the literature \cite{2004-Heilig}, which indicate that the isotope shifts are not as well-understood as claimed and may benefit from further discussion.

Conversely, if relativistic calculations on many electron correlations reach a point where it is possible to predict the specific mass shift for any of the most accurately known neon transitions (i.e. the $640$ nm cooling transition, between levels which have the same core) to an accuracy of roughly $200$ kHz, then $\delta \langle r^2\rangle^{20,22}$ may be deduced directly from the isotope shift without a King plot procedure. This would remove systematic uncertainties in $\delta \langle r^2\rangle^{20,A}$, which are up to an order of magnitude greater than the statistical uncertainty in the IS measurements.
Such an effort is under way.

\subsection{Dual Sideband Spectroscopy}

We measured the isotope shifts of seven transitions with an accuracy of $<10^{-4}$. The consistency of experimental results demonstrates that dual sideband spectroscopy is a simple and precise technique for obtaining the frequency differences between far resonances, and removes frequency noise and nonlinear scanning effects. 

Utilizing this technique a substantially shorter laser frequency scan is possible, which may be beneficial in the spectroscopy of short lived radioisotopes \cite{2016-Pearson}.
We plan to combine this technique with metastable quenching spectroscopy \cite{2015-RFsource}, for hyperfine and isotope shift measurements of various neon isotopes produced at the SARAF particle accelerator \cite{2018-Weak, 2018-Mardor}.

\subsection{Parametric analysis and global fit}

The great body of isotope shift data, which is available only for neon and analyzed in this work, provides a unique opportunity for developing new methods for classifying and understanding the various relativistic, electronic and nuclear effects which donate to the lineshifts.

The parametric investigation employed here, allows to disentangle the complex intermediate coupling of neon levels, and enables to describe $145$ average transition isotope shifts through $21$ effective parameters which are either purely \textit{jj} or LS-coupled. Recalculating neon levelshifts from these parameters results in a substantial increase in precision.
This database of precise neon isotope shifts from $80-8000$ nm may be used for calibration of collinear laser spectroscopy experiments at various accelerator facilities \cite{2015-Schmidt}.

Even though our reanalysis demonstrates that specific mass shift calculations in neon preform better than common knowledge, still their limited accuracy is the limiting factor for most charge radii difference determinations, in neon and other multielectron elements \cite{2015Marinova, 2015-Schmidt}. We propose that new, relativistic \textit{ab initio} mass shift calculations are preformed for a large number of levels, analyzed in the intermediate coupling framework, and compared to experimental observables in a parametric basis.

\begin{acknowledgments}
The work presented here is supported by grants from the Pazy Foundation (Israel), Israel Science Foundation (grants no. 139/15 and 1446/16), the European Research Council (grant no. 714118 TRAPLAB), and the Australian Research Council (DP190100974). BO is supported by the Ministry of Science and Technology, under the Eshkol Fellowship.
\end{acknowledgments}

\clearpage
\appendix

\begin{minipage}{1.0\textwidth}
\section*{Appendix}
This appendix contains the inputs to, and outputs from, our global analysis.
Tables \ref{tab:Lineshifts1} and \ref{tab:Lineshifts2} contain the experimental transition isotope shifts used. Their averaging procedure is discussed in Sec. \ref{chap:Transitions}. Tables  \ref{tab:Level_RIS} and \ref{tab:LevelCorr_2}, present the extracted residual isotope shift (RIS) of levels and their correlations, respectively, relative to the the reference level $3s [3/2]_2$, and discussed in Sec. \ref{chap:Levels}. The angular coefficients, calculated according the procedure outlined in Sec. \ref{chap:IC}, are presented in table \ref{tab:AC}. These coefficients are used for calculation of level shifts for the two fitting procedures outlined in sections \ref{chap:Intra}, and \ref{chap:Inter}.
\end{minipage}

\begin{table}[htbp]
\tiny 

  \caption{Measured isotope shifts in MHz for transitions with air wavelength of 60-600 nm. `Mean' column is the dispersion-corrected weighted average. `Calc' are the calculated shifts from fitting equation \ref{eq:LinesTrans}. }

\begin{tabular}{cccrrrr}
    $\lambda$ & \textbf{From} & \textbf{To} &  \cite{1937-RS,1975-BGG,1979GBC,1994-74nm,2004-Merkt} & \cite{2015Sap,2016-Sap,2017Sap} & $\textbf{Mean IS}$& $\textbf{Calculated IS}$ \\
\\
    62.95  & $^1S_0$        & $4s$ $[3/2]_1$ & 2159(90)     & 2305(20)     & 2298(31)     & 2309(29) \\
    73.57  & $^1S_0$        & $3s$'$[1/2]_1$ & -90(90)      & -98(16)      & -98(16)      & -99(11) \\
    74.35  & $^1S_0$        & $3s$ $[3/2]_1$ & 420(90)      & 417(16)      & 417(16)      & 415(11) \\
    296.07 & $3s$ $[3/2]_2$ & $4d$'$[3/2]_1$ & 2764.6(3.0)  &              & 2764.6(3.0)  & 2764.2(3.0) \\
    296.16 & $3s$ $[3/2]_2$ & $4d$'$[3/2]_2$ & 2779.2(4.0)  &              & 2779.2(4.0)  & 2778.1(3.8) \\
    296.17 & $3s$ $[3/2]_2$ & $4d$'$[5/2]_3$ & 2780.1(1.3)  &              & 2780.1(1.3)  & 2780.2(1.3) \\
\end{tabular}%
    
  \label{tab:Lineshifts1}%

    \begin{tabular}{cccrrrrrrrrrr}
    \\  &   &   & \cite{1930NM,1937-RS,1939-Scho} & \cite{1965Odin} & \cite{1978-WBL} & \cite{1983-BGK,1991-LNS} & \cite{1975-DK,1978-Berg,1988-GDS,1991-Bielski} & \cite{1992-KKR,2000-Cheng}  & \cite{1995-GW} & \cite{1989-SJF,1997Basar} & \textbf{Mean IS} & \textbf{Calculated IS} \\
\\  
  \: 511.4       & $3p$ $[1/2]_1$ & $4d$'$[3/2]_1$ & 899(150)    &             &             &             &             &             &             &             & 899(150)    & 1148.5(3.5) \\
  \: 532.6       & $3p$ $[1/2]_1$ & $4d$ $[3/2]_1$ & 989(90)     &             &             &             &             &             &             &             & 989(90)     & 1052.9(5.2) \\
  \: 533.1       & $3p$ $[1/2]_1$ & $4d$ $[3/2]_2$ & 1229(90)    &             &             &             &             &             &             &             & 1229(90)    & 1082.7(2.8) \\
  \: 534.1       & $3p$ $[1/2]_1$ & $4d$ $[1/2]_1$ & 1139(90)    &             &             &             &             &             &             &             & 1139(90)    & 1070.3(3.2) \\
  \: 534.3       & $3p$ $[1/2]_1$ & $4d$ $[1/2]_0$ & 929(90)     &             &             &             &             &             &             &             & 929(90)     & 929(90) \\
\end{tabular}%
\end{table}%

\clearpage

\begin{table}[htbp]
\tiny
\centering
  \caption{Measured isotope shifts in MHz for transitions with air wavelength of 600-10,000 nm. `Mean' column is the dispersion-corrected weighted average. `Calculated' are the calculated shifts from fitting equation \ref{eq:LinesTrans}. Ouliers are italicized.}
    \begin{tabular}{cccrrrrrrrrrr}
  $\lambda $ & \textbf{From} & \textbf{To}  & \cite{1937-RS,1939-Scho,1965-Cor}& \cite{1965Odin,1966-Zory}& \cite{1968-Schw,1975-DK,1980-JPL,1983-BGK}  & \cite{1989-SJF,1991-LNS,1992-KKR} & \cite{1972-Balik,1995-GW}  & \cite{1980-Kotlikov,1997Basar} & \cite{1967-VS,1972-DFR,1987-Zhao,2008Mass} &\textbf{This Work} & \textbf{Mean} & \textbf{Calculated} \\\\
    600.1  & $3p$ $[3/2]_2$ & $4d$ $[1/2]_1$ &               &               &               &          		 &               & 1020(20)     		  &               	  &               & 1020(20)      & 1022.0(2.6) \\
    603.0  & $3s$ $[3/2]_1$ & $3p$'$[1/2]_1$ & 1709(90)      &               &               &          		 & 1743.5(5.0)   & 1746(17)     		  &               	  &               & 1743.6(4.8)   & 1747.18(53) \\
    607.4  & $3s$ $[3/2]_1$ & $3p$ $[1/2]_0$ & 1679(90)      & 1702.8(3.0)   &               &             		 & 1700.5(5.0)   & 1695(17)     		  &               	  &               & 1702.0(2.5)   & 1701.6(2.5) \\
    609.6  & $3s$ $[3/2]_1$ & $3p$'$[3/2]_2$ & 1679(90)      & 1726.8(3.0)   & 1740(12)      & \textit{1753(8)}  & 1720.0(5.0)   & 1722(17)     		  & 1708(26)      	  &               & 1725.3(2.5)   & 1726.03(95) \\
    612.8  & $3s$ $[3/2]_1$ & $3p$'$[3/2]_1$ &               &               &               &              	 & 1737.5(2.0)   & 1745(16)     		  &               	  &               & 1737.6(2.0)   & 1739.04(25) \\
    614.3  & $3s$ $[3/2]_2$ & $3p$ $[3/2]_2$ & 1649(90)      & 1660.9(3.0)   & 1669(4)  	 & 1660(3)  		 & 1663.7(5.0)   & 1653(14)     		  & 1663.6(1.7)   	  &               & 1663.0(1.2)   & 1664.01(25) \\
\end{tabular}%
  \label{tab:Lineshifts2}%

    \begin{tabular}{cccrrrrrrr}
  \\&   &    & \cite{1937-RS,1939-Scho,1963-Szoke,1966-Sak} & \cite{1967-Cor} & \cite{1972-DFR} & \cite{1972-Balik,1973-Ohi} & \cite{1973Keller,1975-DK,1992-SHK}  & \textbf{Mean} & \textbf{Calculated} \\\\
813.6        & $3p$ $[3/2]_1$         & $3d$'$[5/2]_2$         & 570(150)               &            &                        &                        &                        & 570(150)               & 722.1(6.4) \\
    837.7        & $3p$ $[5/2]_3$         & $3d$ $[7/2]_4$         & 630(150)               &            &                        &                        &                        & 630(150)               & 714(77) \\
    849.5        & $3p$ $[5/2]_2$         & $3d$ $[7/2]_3$         & 630(90)                &            &                        &                        &                        & 630(90)                & 657.4(6.7) \\
    859.1        & $3p$'$[3/2]_1$         & $3d$'$[5/2]_2$         & 510(150)               &            &                        &                        &                        & 510(150)               & 662.8(6.4) \\
    863.4        & $3p$ $[3/2]_1$         & $3d$ $[5/2]_2$         & 600(150)               &            &                        &                        &                        & 600(150)               & 618(47) \\
\end{tabular}%

\end{table}%

\clearpage

\begin{table}[!htb]
  \centering
\scriptsize
  \caption{Residual isotope shifts in MHz, relative to the reference level. The \nth{1} column is obtained from fitting Eq. \ref{eq:LinesTrans}, and the last two columns by fitting Eq. \ref{eq:ParTrans} utilizing the models of Sec. \ref{chap:Intra} and \ref{chap:Inter}.
  }
  \begin{ruledtabular}
    \begin{tabular}{crrr}
                    & \multicolumn{1}{c}{\textbf{Direct calculation} } & \textbf{Intraconf. fit} & \textbf{Interconf. fit} \\
    \textbf{Parameters:} &\multicolumn{1}{c}{55} & \multicolumn{1}{c}{30} & \multicolumn{1}{c}{21} \\
    \textbf{DOF}   & \multicolumn{1}{c}{90} & \multicolumn{1}{c}{115} & \multicolumn{1}{c}{124} \\
    \textbf{Chi2/DOF} & \multicolumn{1}{c}{0.987} & \multicolumn{1}{c}{0.937} & \multicolumn{1}{c}{0.930} \\
    \textbf{Relative AIC} & \multicolumn{1}{c}{42} & \multicolumn{1}{c}{10} & \multicolumn{1}{c}{0} \\\\
	\cline{1-4}  
    $S_0$           & 9597(11)        & 9598(11)        & 9602(11) \\
    \cline{1-4}  
    $3s$ $[3/2]_2$  & \multicolumn{1}{c}{\textbf{0}}   & \multicolumn{1}{c}{\textbf{0}}       & \multicolumn{1}{c}{\textbf{0}}  \\
\end{tabular}%
    \end{ruledtabular}
  \label{tab:Level_RIS}%
\end{table}%

\clearpage
\begin{turnpage}
\begin{table}[!ht]
  \centering
  \tiny
  \caption{Correlation matrix of levels calculated using Eq. \ref{eq:LinesTrans}.}
    \begin{tabular}{cccccccccccccc}
                         & $S_0$                & $3s$ $[3/2]_1$       & $3s$'$[1/2]_0$      & $3s$'$[1/2]_1$      & $3p$ $[5/2]_2$       & $3p$ $[3/2]_1$       & $3p$ $[3/2]_2$       & $3p$'$[3/2]_1$      & $3p$'$[3/2]_2$      & $3p$'$[1/2]_1$      & $3p$'$[1/2]_0$      & $4s$ $[3/2]_1$      & $4s$'$[1/2]_1$ \\
    $S_0$                & 1                    &                      &                      &                      &                      &                      &                      &                      &                      &                      &                      & 0.33                 &  \\
    $3s$ $[3/2]_1$       &                      & 1                    &                      &                      & 0.46                 & 0.19                 & 0.65                 &                      &                      &                      &                      &                      &  \\
    $3s$'$[1/2]_0$       &                      &                      & 1                    &                      &                      & 0.43                 &                      & 0.86                 &                      &                      &                      &                      &  \\
    $3s$'$[1/2]_1$       &                      &                      &                      & 1                    &                      &                      &                      &                      & 0.12                 & 0.20                 & 0.37                 &                      & 0.18 \\
    $3p$ $[5/2]_2$       &                      & 0.46                 &                      &                      & 1                    &                      & 0.30                 &                      &                      &                      &                      &                      &  \\
\end{tabular}%

    \begin{tabular}{cccccccccccccccccccccc}
                           & $3d$ $[7/2]_4$         & $3d$ $[7/2]_3$         & $3d$ $[3/2]_2$         & $3d$ $[3/2]_1$         & $3d$ $[5/2]_2$         & $3d$ $[5/2]_3$         & $3d$'$[5/2]_2$         & $3d$'$[5/2]_3$         & $3d$'$[3/2]_2$         & $3d$'$[3/2]_1$         & $4p$ $[5/2]_3$         & $4p$ $[5/2]_2$         & $4p$ $[3/2]_1$         & $4p$ $[3/2]_2$         & $4p$ $[1/2]_0$         & $4p$'$[3/2]_1$         & $4p$'$[1/2]_1$         & $4p$'$[3/2]_2$         & $4p$'$[1/2]_0$         & $5s$ $[3/2]_1$         & $5s$'$[1/2]_1$ \\
    $3d$ $[7/2]_4$         & 1                      &                        &                        &                        &                        &                        &                        &                        &                        &                        & 0.999970               &                        &                        &                        &                        &                        &                        &                        &                        &                        &  \\
    $3d$ $[7/2]_3$         &                        & 1                      &                        &                        &                        &                        &                        &                        &                        &                        &                        & 0.9909                 &                        &                        &                        &                        &                        &                        &                        & 0.44                   &  \\
\end{tabular}%
  \label{tab:LevelCorr_2}%
\end{table}%
\end{turnpage}

\clearpage
\begin{table}[htbp]
  \centering
  \scriptsize
  \caption{
  Parentage of neon levels in percentage. The first six columns are the squared angular coefficients in an LS-coupled basis. The last two columns are the coefficients in a \textit{jj}-coupled basis. On top are odd levels and on bottom are even levels.
  }
    \begin{tabular}{lcccccccc}\\
                              &  $^1P$             & $^3P$              &  $^1D$               & $^3D$               &  $^1F$             & $^3F$              &$j_\mathrm{c}=3/2$        & $j_\mathrm{c}=1/2$    \\

   \cline{1-9}   $3s$ $[3/2]_2$            &                           & 100                      &                           &                           &                           &                           & 100                      & \\
    		            $3s$ $[3/2]_1$            & 7.06                     & 92.94                    &                           &                           &                           &                           & 59.8                     & 40.2  \\
  		              $3s$'$[1/2]_0$            &                           & 100                      &                           &                           &                           &                           &                           & 100  \\
 		               $3s$'$[1/2]_1$            & 92.94                    & 7.06                     &                           &                           &                           &                           & 40.16                    & 59.84   \\

  \cline{1-9}  $4s$ $[3/2]_1$            &                           & 100                      &                           &                           &                           &                           & 100.00                   &   \\
					   $4s$ $[3/2]_1$            & 43.90                    & 56.10                    &                           &                           &                           &                           & 94.75                    & 5.25   \\
\end{tabular}%
  \label{tab:AC}%

    \begin{tabular}{lcccccccc}\\
 & $^1S$             & $^3S$      &  $^1P$             & $^3P$              &  $^1D$               & $^3D$              &$j_\mathrm{c}=3/2$        & $j_\mathrm{c}=1/2$    \\

\cline{1-9} 
\cline{1-9}    $3p$ $[1/2]_1$            &                           & 98.22                    & 0.57                     & 1.22                     &                           & 0.00                     & 78.55                    & 21.45   \\
						  $3p$ $[5/2]_3$            &                           &                           &                           &                           &                           & 100                      & 100                      &   \\
						  $3p$ $[5/2]_2$            &                           &                           &                           & 2.08                     & 22.62                    & 75.31                    & 92.15                    & 7.85   \\
						  $3p$ $[3/2]_1$            &                           & 0.001                    & 17.687                   & 9.147                    &                           & 73.166                   & 67.58                    & 32.42   \\
						  $3p$ $[3/2]_2$            &                           &                           &                           & 43.42                    & 49.05                    & 7.53                     & 89.15                    & 10.85   \\
						  $3p$'$[3/2]_1$            &                           & 0.00                     & 49.14                    & 24.03                    &                           & 26.83                    & 32.42                    & 67.58   \\
\end{tabular}%
  \label{tab:AC2}%
\end{table}%

\clearpage

\bibliography{sample.bib}

\end{document}